\documentclass[epj]{svjour}
\usepackage{graphics}
\begin{document}
\title{Non-empirical pairing energy density functional}
\subtitle{First order in the nuclear plus Coulomb two-body interaction}
\author{T. Lesinski\inst{1}\thanks{\emph{Present address:}Physics Division, Oak Ridge National
Laboratory, Oak Ridge, TN 37831, USA} \and T. Duguet\inst{2,3} \and K. Bennaceur \inst{1} \and J. Meyer \inst{1}
}                     
\institute{Universit{\'e} de Lyon, F-69003 Lyon, France;
             Universit{\'e} Lyon 1, F-69622 Villeurbanne, France;
             CNRS/IN2P3; Institut de Physique Nucl{\'e}aire de Lyon \and CEA, Centre de Saclay, IRFU/Service de Physique Nucl{\'e}aire, F-91191 Gif-sur-Yvette, France \and National Superconducting Cyclotron Laboratory
             and Department of Physics and Astronomy,
             Michigan State University, East Lansing, MI 48824, USA}
%
%
\abstract{We perform systematic calculations of pairing gaps in semi-magic nuclei across the nuclear chart using the Energy Density Functional method and a {\it non-empirical} pairing functional derived, without further approximation, at lowest order in the two-nucleon vacuum interaction, including the Coulomb force. The correlated single-particle motion is accounted for by the SLy4 semi-empirical functional. Rather unexpectedly, both neutron and proton pairing gaps thus generated are systematically close to experimental data. Such a result further suggests that missing effects, i.e. higher partial-waves of the NN interaction, the NNN interaction and the coupling to collective fluctuations, provide an overall contribution that is sub-leading as for generating pairing gaps in nuclei. We find that including the Coulomb interaction is essential as it reduces proton pairing gaps by up to  $40\%$.
} 
\maketitle
\section{Introduction}
\label{intro}

It was realized long ago~\cite{Bohr58} that features such as the odd-even staggering of nuclear binding energies, or moments of inertia having half the corresponding rigid-body value, were related to the superfluid character of atomic nuclei. As a matter of fact, like-particle pairing has become an essential ingredient of nuclear-structure models, in particular regarding the description of exotic nuclei~\cite{doba03a}. In addition, superfluidity plays a key role in neutron stars, e.g. it strongly impacts post-glitch timing observations~\cite{Avogadro07} or their cooling history~\cite{heiselberg3}.

Although {\it ab-initio} calculations of pairing properties are possible for the idealized infinite-nuclear-matter system~\cite{Dean03,Baldo07,Hebeler07}, they remain a challenge for finite nuclei beyond the lightest ones. In particular, finite-size effects require to go beyond a simple extrapolation of the results obtained in infinite nuclear matter. The nuclear Energy Density Functional (EDF) approach is the microscopic tool of choice to study medium-mass and heavy nuclei in a systematic manner~\cite{bender03a}. Within a single-reference (SR) implementation, pairing is treated through $U(1)$ symmetry breaking which leads to solving Hartree-Fock-Bogoliubov (HFB)~\cite{RingSchuck} or Bogoliubov-de Gennes equations. However, the latter are usually solved in terms of empirical nuclear functionals for both single-particle and pairing channels. It is of current interest to construct {\it non-empirical} energy density functionals derived explicitly from two- and three-nucleon vacuum interactions~\cite{Duguet06}, e.g. on the basis of many-body perturbation theory (MBPT), in view of the challenge posed by exotic nuclei displaying an unusually large ratio of neutrons over protons. The recent advent of low-momentum nuclear interactions based on renormalization group (RG) techniques~\cite{Bogner03} opens up such a possibility.

Describing pairing through MBPT \cite{Abrikosov,Nozieres} translates into solving a generalized Bardeen-Cooper-Schrieffer gap equation. Doing so requires two essential ingredients: the normal self-energy function, summing interaction processes between a single nucleon and the medium, together with the pairing interaction kernel, both of which can be expanded in terms of the vacuum interaction vertex. The
first-order contribution to the two-particle irreducible pairing kernel is the vacuum interaction itself, while higher-order terms
include the induced interaction describing the process of paired particles interacting
through the exchange of collective medium fluctuations. A fundamental, yet unresolved, question relates to how much of the pairing gaps in finite nuclei are accounted for at lowest order, i.e. by the direct term of the two-nucleon interaction, and how much is due to higher-order processes. That the direct nuclear interaction can generate superfluidity is at variance with the situation in electronic systems.

Calculations of pairing gaps in a non-homogeneous system have been performed for a slab of nuclear matter~\cite{Baldo03,Pankratov07} building the pairing kernel at lowest order in the vacuum interaction and complementing it with a schematic normal self-energy. Rare studies performed in finite nuclei tend to show that the direct term of the vacuum interaction only accounts for a fraction of experimental
pairing gaps~\cite{Barranco04}, and that induced interactions due to coupling with collective modes~\cite{Terasaki02,Giovanardi02} can explain the remainder~\cite{Barranco04,Gori05,Pastore08}. However, due to the complexity of such calculations, only a single nucleus ($^{120}${Sn}) has been studied.

In the present paper, we contribute to the discussion by performing a systematic study of gaps generated by a pairing interaction kernel computed at first order in the two-body (NN) vacuum interaction, using the method outlined below~\cite{Duguet07,lesinski07a} and detailed in a forthcoming publication~\cite{Lesinski08}. We shall treat the NN interaction fully, including hadronic charge-symmetry-breaking terms and the Coulomb interaction, but shall omit the three-nucleon (NNN) interaction at this point.

\section{Method}
\label{sec:1}

We use the low-momentum NN interaction V$_{{\rm low \, k}}$~\cite{Bogner03} built from the Argonne $v_{18}$
NN potential~\cite{Wiringa95} with a renormalization cut-off $\Lambda=2.5$\,fm$^{-1}$. As for the description of low-energy nuclear phenomena, V$_{{\rm low \, k}}$ is as a valid {\it vacuum} NN interaction as any more conventional NN potential that tries to model explicitly the short-range, i.e. high-energy physics. As a result, pairing gaps computed in infinite nuclear matter (INM) using free kinetic energy and a pairing kernel at lowest order in the NN interaction are essentially independent of the intrinsic resolution scale $\Lambda$ that characterizes the latter, from $\Lambda=1.8$ fm$^{-1}$ up to values that are typical of the Argonne $v_{18}$ potential~\cite{Hebeler07}.

Low-momentum interactions are technically easier to handle in many-body calculations than conventional hard-core potentials. In particular, low-momentum interactions are easily amenable to a low-rank separable representation \cite{Bogner06b} which
enables their efficient use in a dedicated EDF code~\cite{Duguet07,Duguet04}. Nuclear superfluidity at sub-saturation densities is mostly generated in the spin-singlet and isospin-triplet channels of the $L=0$ partial wave of relative
motion~\cite{Dean03}. We thus generate a high-precision rank-three separable representation of the numerical V$_{{\rm low \, k}}$ in the $^1S_0$ channel~\cite{lesinski07a}, whose spatial part reads
\begin{eqnarray}
\langle \vec{r}_1 \vec{r}_2 \vert {\rm V}^{S} \vert \vec{r}_3 \vec{r}_4 \rangle
    &=& \! \sum_{\alpha=1}^3 ~
        \lambda_{\alpha}~ G_\alpha(s_{12})~ G_\alpha(s_{34})
            \,   \delta(\vec{R}_{12}\!-\!\vec{R}_{34}) \, ,
    \label{eq:sep:matelemr}
\end{eqnarray}
where relative and center-of-mass coordinates are defined as $s_{ij}=|\vec{s}_{ij}| =
|\vec{r}_i - \vec{r}_j|$ and $\vec{R}_{ij} = (\vec{r}_i + \vec{r}_j)/2$, respectively. In Eq.~\ref{eq:sep:matelemr}, $G_\alpha(s)$ denotes coordinate-space form factors whose analytical expression and free parameters are optimized to reproduce the nuclear part of V$_{{\rm low \, k}}$ matrix elements and the corresponding scattering phase-shifts. As for the Coulomb contribution to the proton-proton pairing kernel, an accurate separable representation of the Coulomb interaction truncated at a radius $r=2R_{\rm nucleus}$ is built~\cite{lesinski07a}. Having a separable representation of the NN interaction is essential to obtain a non-local pairing functional leading to HFB equations that are of a similar computational burden to those generated by local (empirical) pairing functionals~\cite{lesinski07a}.

The remaining part of the nuclear EDF, i.e. the part accounting for correlated single-particle motions, is taken as the semi-empirical SLy4 Skyrme functional~\cite{chabanat98}. The corresponding density-dependent but momentum-independent isoscalar
effective-mass is equal to $m^{\ast}_{0}=0.7 \, m$ at nuclear saturation density which is consistent with Hartree-Fock results obtained with low-momentum interactions~\cite{KaiAchimTD}. Consistently with the pairing vertex used, such a value of $m^{\ast}_{0}$ corresponds to leaving out contributions to the normal self-energy associated with the coupling to collective fluctuations. Higher-order contributions to the pairing kernel and the normal self-energy are left out for future works.

The BSLHFB code~\cite{lesinski08b} solves HFB equations in a spherical box of 24\,fm radius, with a mesh step
of 0.3\,fm. Single-particle wave functions are expanded on a basis of spherical
Bessel functions $j_\ell(kr)$ with a momentum cut-off $k_{\rm cut} =
4.0$\,fm$^{-1}$ allowing the description of single-particle states up to energies
of about 300\,MeV and ensuring convergence of the pairing gaps to a fraction of a keV.

\section{Results}
\label{sec:2}

\begin{figure*}
\resizebox{\textwidth}{!}{%
  \includegraphics{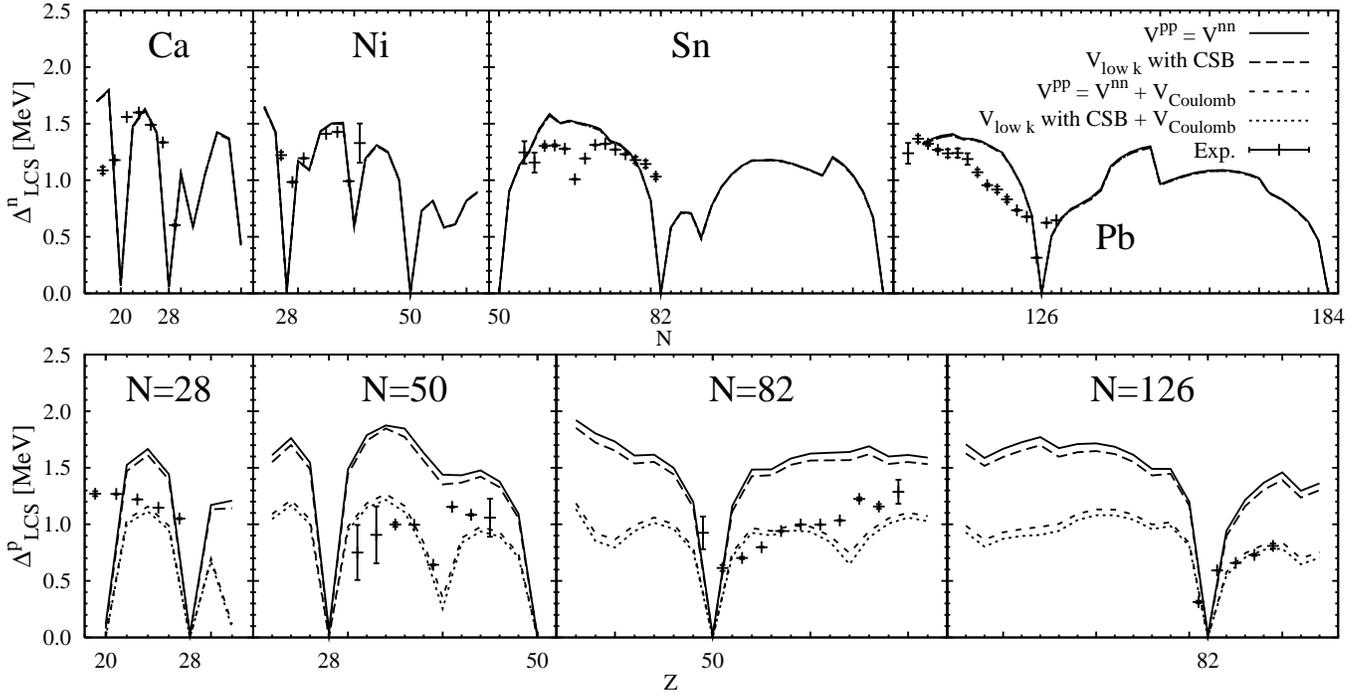}
}
\caption{Neutron/proton gaps along isotopic/isotonic chains obtained using the neutron-neutron part of V$_{{\rm low \, k}}$ for both the neutron and proton pairing kernels, the full charge-symmetry breaking (CSB)
    nuclear V$_{{\rm low \, k}}$ while including or excluding the Coulomb interaction.}
\label{gaps:coulomb}
\end{figure*}

We presently limit ourselves to discussing a single observable related to pairing correlations, i.e. the
odd-even mass staggering (OEMS), whereas additional results will be reported in a separate publication~\cite{Lesinski08}. The connection between finite-difference mass formulae employed to extract the OEMS and
theoretical pairing gaps is not straightforward. In the case of
strong pairing and tightly-spaced single-particle levels, the experimental
three-point mass difference formula $\Delta^{(3)}_{q}(N/Z)$ centered on odd $N/Z$ provides the best estimate of
theoretical pairing gaps calculated in even-even nuclei~\cite{Duguet01}. The latter theoretical gaps are provided by
$\Delta^q_{\rm LCS}$ which denotes the diagonal pairing matrix
element $\Delta^q_k$ corresponding to the \emph{canonical} single-particle state
$\phi_k$ whose quasi-particle energy\footnote{The acronym {\it LCS} stands for {\it Lowest Canonical State}.}
\begin{eqnarray}
 E^q_k &=& \sqrt{(\varepsilon^q_k - \lambda^q)^2 + \Delta_k^{q \, 2}},
\end{eqnarray}
is the lowest, where $\varepsilon^q_k$ is the diagonal matrix element of the single-particle field $h^q$ in $\phi_k$, $\lambda^q$ the chemical potential for the species of isospin projection $q$ and $\Delta^q_k$ the corresponding diagonal pairing-field matrix element.

Figure~\ref{gaps:coulomb} displays neutron/proton gaps $\Delta^{{\rm n/p}}_{\rm LCS}$ along all semi-magic isotopic/isotonic chains from proton to neutron drip lines, together with experimental values of $\Delta^{(3)}_{q}$ where available. Theoretical pairing gaps are zero at (sub-) shell closures. Moving away from the latter, they increase more slowly than data. It is known that variation after particle-number restoration, or an approximate variant thereof such as the Lipkin-Nogami method, increase gaps near shell closures~\cite{Stoitsov07}. It is thus necessary for such nuclei to perform a more involved multi-reference (MR) EDF calculation, which is beyond the scope of the present work.

Let us first focus on neutron gaps in the upper panel of Fig.~\ref{gaps:coulomb}. The main message to convey is the overall closeness of the predictions with experimental data; this being true from rather light nuclei (calcium) up to heavy ones (lead). This comes as a surprise given that the calculation is performed at first-order in the NN interaction and that only two thirds of the experimental neutron pairing gap in $^{120}$Sn was obtained in Ref.~\cite{Pastore08} through a calculation similar to the one performed here for many more nuclei. Also, it was advocated there that the missing one-third was provided by the coupling to collective fluctuations. We will briefly comment on the mismatch between the two sets of calculations in the last section of the present paper.

On a detailed basis, of course, the agreement between theoretical predictions and experimental data is not perfect. In addition to the fact that no agreement is to be expected at the present level of many-body treatment, why one should not aim at a detailed, nucleus-by-nucleus, analysis at this point because (i) comparing $\Delta^q_{\rm LCS}$ and $\Delta^{(3)}_{q}(N/Z)$ is only an appropriate zeroth-order procedure~\cite{Duguet01}, (ii) gaps in a particular nucleus or region may be spoiled by the semi-empirical nature of the Skyrme EDF that generates the underlying single-particle spectrum; e.g. the depletion of gaps around $N=65$ in tin suggests the existence of a sub-shell closure not predicted by SLy4 whereas the decrease before $N=126$ in lead isotopes is steeper for the calculated gaps, suggesting too-large a level density consistently with the too high-lying $\nu1i_{13/2}$ single-particle shell~\cite{Lesinski07}.

Let us now come to proton gaps along semi-magic isotonic chains. Calculating proton gaps requires to incorporate charge-symmetry breaking (CSB) contributions to the NN interaction, which are of two distinct origins (i) electromagnetic, principally through the Coulomb interaction between protons (ii) hadronic, more slightly breaking the symmetry such that the NN interaction between protons is less attractive than between neutrons. Four sets of results for proton gaps are displayed in the lower panel of Fig.~\ref{gaps:coulomb} that correspond to incorporating/omitting hadronic and/or Coulomb CSB terms in the NN interaction that builds the proton pairing interaction kernel.

Proton gaps generated from a purely charge-symmetric pairing functional over-estimate experimental data systematically. As a matter of fact, calculated proton gaps are, in the heaviest isotonic chains, larger than neutron gaps in neighboring isotopic chains, with values standing above 1.5\,MeV for protons and between 1 and 1.5\,MeV for neutrons. Although it is known that proton gaps are similar in
magnitude, or marginally larger, than neutron ones in heavy nuclei~\cite{Nemirovsky62}, the difference observed here is larger than the
one present in experimental data. Constructing the pairing kernel from the charge-symmetric part of V$_{{\rm low \, k}}$, larger proton gaps may be traced back to the neutron excess in heavy nuclei which makes the proton density and the local proton Fermi momentum to be lower than the
neutron ones. As a result, the momentum density distribution
of proton states close to the Fermi energy is peaked at lower
momenta than for neutrons which makes the proton pairing tensor to probe more attractive matrix
elements of the NN interaction than the neutron one; see Ref.~\cite{Bogner03} for $V_{{\rm low\,k}}(k,k')$ matrix elements in the $^{1}S_{0}$ channel. The same effect can be invoked to some extent for the
neutron-excess-dependence of gaps: $\Delta^{{\rm n}}_{\rm LCS}$ globally decreases with $N$ for all four chains shown in Fig.~\ref{gaps:coulomb}, due
to the increase of the neutron density and the local Fermi momentum.
Proton gaps exhibit a less marked decrease with $Z$,
probably attributable to the Coulomb barrier in the single-particle proton field. As discussed in Ref.~\cite{Lesinski08}, such isospin trends are partly washed out when the wrong isovector effective mass of the SLy4 functional is corrected~\cite{lesinski06a}.

Including hadronic CSB contributions only produces a slight shift of proton-gap curves which is negligible compared to Coulomb effects. Coulomb decreases $\Delta^p_{\rm LCS}$ values by $30\%$ to $50\%$. Once it is included, an unexpected and rather impressive agreement with experiment is obtained for both neutron and proton gaps. Again, a detailed comparison reveals qualitative discrepancies that can often be attributed to the underlying single-particle spectrum generated by the Skyrme semi-empirical functional; e.g. in
$N=50$ isotones, the relative magnitude of gaps below and above $Z=40$ is not captured, while the decrease of the gap at $Z=40$ is quite stronger than in data, which do not clearly distinguish between a shell closure at $_{38}$Sr or at $_{40}$Zr. This points to the inappropriate predictions of level spacings by SLy4 in this region, the position of the $1g_{9/2}$ shell
being too high.

We are aware of only one other systematic HFB calculation explicitly including the Coulomb
interaction in the proton pairing channel~\cite{Anguiano01}. It was performed in a
triaxial harmonic-oscillator basis using the Gogny D1 or D1S semi-empirical kernels in both channels of the EDF. Although
no specific study of pairing gaps was proposed in Ref.~\cite{Anguiano01}, it was found that proton
pairing energies were reduced by 30 to as much as 60\% (for semi-magic
$^{90}${Zr}) when including the Coulomb pairing term self-consistently in the
variational procedure. Lowest two-quasiproton energies, which are the quantities discussed in Ref.\,\cite{Anguiano01} that relate the most to pairing gaps, were reduced by 20 to 30\%. Bearing in mind the different EDF used for the hadronic part, the reduction of the pairing gaps observed in our study agrees with the conclusions of Ref.\,\cite{Anguiano01}. Such a reduction due to the Coulomb interaction is large enough to be systematically taken into consideration in HFB calculations. When using empirical local pairing functionals, it validates the use of distinct values for neutron and proton pairing parameters~\cite{Goriely06}, although additional (effective) effects come into play in this context~\cite{yamagami08a}.

\section{Discussion, conclusions and outlook}
\label{sec:3}

A key result of the present work is that proton and neutron pairing gaps computed at lowest order in the low-momentum nuclear plus Coulomb two-body interaction are close to experimental data for a large set of semi-magic spherical nuclei. Such a result, which was neither expected nor hoped for, remains unchanged~\cite{Lesinski08} over the whole range of renormalization group cut-off values $\Lambda \in \left[ 1.8, 3.0 \right]$ fm$^{-1}$ that characterizes perturbative NN interactions~\cite{Bogner06b,Bogner05}. Also, such a result does not depend significantly on the Skyrme isoscalar effective mass~\cite{Lesinski08}, as long as the value of the latter at saturation density is within the interval $[0.65,0.75]$ authorized by ab-initio calculations of Infinite Nuclear Matter consistent with the use of the vacuum NN interaction in the pairing channel~\cite{KaiAchimTD}. The closeness of presently computed pairing gaps with experimental data does raise legitimate questions, some of which are briefly addressed below and more extensively in Refs.~\cite{Lesinski08,KaiAchimTD}. In any case, it is clear that the situation in nuclei is largely different from the one encountered in electronic systems where the direct electron-electron interaction cannot contribute to building superfluidity.

Of course, it is essential to check that computing the normal part of the first-order self-energy directly from low-momentum NN interactions, rather than from the semi-empirical SLy4 functional, does not alter the results reported here. It is shown in Ref.~\cite{KaiAchimTD} that pairing gaps obtained in infinite nuclear matter using both types of normal self-energy are the same over the density range that is relevant to finite nuclei. Still, such an equivalence remains to be studied directly in finite nuclei by computing both the normal~\cite{coraggio03a,roth06a} and anomalous first-order self-energies from the same low-momentum interaction\footnote{References~\cite{coraggio03a,roth06a} focus on the Hartree-Fock single-particle spectrum of $^{40}$Ca which displays significant differences with the corresponding spectrum generated by Skyrme EDFs characterized by $m^{\ast}_{0}=0.7 \, m$ at nuclear saturation density. However, it should be remembered that $^{40}$Ca constitutes an anomaly as for the spectrum generated by Skyrme EDFs which is unnaturally dense around the Fermi energy~\cite{Lesinski07,brown98a}.  In addition, and even though the density of states should be mainly governed by the NN interaction, a meaningful calculation of single-particle spectra based on microscopic interactions must include the NNN interaction as the latter is expected to contribute significantly to spin-orbit splittings.}.

Keeping such a caveat in mind, the present results indicate that missing effects, i.e. contributions of higher partial-waves of the NN interaction and of the NNN interaction to the spin-singlet/isospin-triplet pairing kernel, as well as the coupling to density, spin and isospin collective fluctuations that come in at higher orders\footnote{We denote here the consistent renormalization of {\it both} the normal self-energy and the pairing interaction kernel through the coupling to collective fluctuations.}, provide an {\it overall} contribution to pairing gaps in nuclei that is sub-leading\footnote{Of course, missing contributions do not have to be individually small or negligible.}. In this context, it remains to be understood how such missing contributions could alter pairing gaps, i.e. the {\it difference} of two successive one-nucleon separation energies, only mildly while coupling to fluctuations are known to strongly renormalize the density of states at the Fermi energy, e.g. the {\it sum} of two successive one-nucleon separation energies. It is thus of great interest in the near future (i) to extend the present study to deformed nuclei, which is underway, and (ii) to check the impact of all missing contributions by incorporating them explicitly, and consistently, in the calculation.

The results presented here are at variance with those of Ref.~\cite{Pastore08} where only two thirds of the experimental neutron pairing gap was obtained in $^{120}$Sn from a calculation similar to the one performed here, except that the {\it hard-core} vacuum Argonne $v_{14}$ NN interaction~\cite{Wiringa95} was used to build the pairing interaction kernel. From a general standpoint, it is first essential to realize that many-body expansion schemes crucially depend on the softness of the NN and NNN interactions, such that finite-order results for soft and hard interactions are not equivalent and immediately comparable. Soft interactions ($\Lambda \leq 3.0$ fm$^{-1}$) rely on a {\it perturbative} expansion whereas hard ones ($\Lambda \geq 3.0$ fm$^{-1}$) require to recast the expansion in terms of a {\it hole-line} expansion. Due to the necessity to rearrange the expansion scheme as one raises the resolution scale $\Lambda$, results obtained through truncated calculations cannot be expected to be independent of that cut-off $\Lambda$. Beyond such a crucial remark, whose consequences for the computation of pairing gaps are extensively addressed in Ref.~\cite{KaiAchimTD}, the difference between our result and the one obtained in Ref.~\cite{Pastore08} relates to another topic; i.e. the inappropriate use of a Skyrme EDF to generate the single-particle field when it is combined with a hard interaction in the pairing channel. As demonstrated in Ref.~\cite{KaiAchimTD}, the momentum-averaging procedure of the effective mass in the vicinity of the Fermi surface
that underlines such a calculation is inappropriate when performing the many-body calculation on the basis of an interaction presenting a large intrinsic momentum resolution scale, as is the case of Argonne $v_{14}$. As such a NN interaction scatters pairs up to momenta about which the
non-approximate momentum-{\it dependent} effective mass has significantly increased beyond its value
at $k \sim k^q_F$, the use of the momentum-{\it independent} Skyrme effective mass leads to an artificial decrease of pairing gaps that accounts for the difference between our result and the one of Ref.~\cite{Pastore08}.

Last but not least, the pairing gap obtained presently for $^{120}$Sn differs from the one computed in Ref.~\cite{Barranco05} using the same low-momentum interaction as in the pairing channel. As demonstrated in Ref.~\cite{Lesinski08}, the results of Ref.~\cite{Barranco05} are plagued with the use of a Woods-Saxon potential characterized by an effective-mass $m^{\ast}_{0}=0.7 \, m$. In addition to being artificially reduced at high momenta, such an effective mass is also position {\it independent}, i.e. it keeps a constant reduced value from inside to outside the nucleus where it should recover its bare value, as is the case for the Skyrme functional. This leads to an artificial reduction of the density of single-particle states around the Fermi energy compared to a Skyrme functional characterized by $m^{\ast}_{0}=0.7 \, m$ at nuclear saturation density. Eventually, this produces an artificial reduction of the pairing gaps that explains, in a quantitative manner, the discrepancy with our results. Note that such an artificial reduction of the density of states around the Fermi energy impacts pairing gaps independently on whether one uses a hard or a soft interaction in the pairing channel.

\section*{Acknowledgments}

This work was supported by the U.S. National Science Foundation under Grant No. PHY-0456903. K.~B. and T.~L. wish to thank the NSCL for its hospitality and support.



\begin{thebibliography}{}
%
%

\bibitem{Bohr58}
A. Bohr, B. R. Mottelson, D. and Pines, Phys. Rev.  \textbf{110} (1958) 936.

\bibitem{doba03a}
J. Dobaczewski, W. Nazarewicz, Prog. Theor. Phys. Suppl. \textbf{146} (2003) 70.

\bibitem{Avogadro07}
P. Avogadro, F. Barranco, R. A. Broglia, E. Vigezzi, Phys. Rev. C \textbf{75}(2007) 012805.

\bibitem{heiselberg3}
H. Heiselberg, M. Hjorth-Jensen, Phys. Rep. \textbf{328} (2000) 237.

\bibitem{Dean03}
D.~J. Dean, M. Hjorth-Jensen, Rev. Mod. Phys. \textbf{75} (2003) 607.

\bibitem{Baldo07}
M. Baldo, H.~J. Schulze, Phys. Rev. C \textbf{75} (2007) 025802.

\bibitem{Hebeler07}
K. Hebeler, A. Schwenk, B. Friman, Phys. Lett. B \textbf{648} (2007) 176.

\bibitem{bender03a}
M. Bender, P.-H. Heenen, P.-G. Reinhard, Rev. Mod. Phys. \textbf{75} (2003) 121.

\bibitem{RingSchuck}
P. Ring, P. Schuck,  \emph{The Nuclear Many-Body Problem}, Springer, Berlin, Heidelberg, (2000).

\bibitem{Duguet06}
T. Duguet, K. Bennaceur, T. Lesinski, J. Meyer, INT proceedings \textbf{15} (2006) 21;
  arXiv:nucl-th/0606037.

\bibitem{Bogner03}
S. K. Bogner, T. T. S. Kuo, A. Schwenk, Phys. Rep. \textbf{386} (2003) 1.

\bibitem{Wiringa95}
R. B. Wiringa, V. G. J. Stoks, R. Schiavilla, Phys. Rev. C \textbf{51} (1995)  38.

\bibitem{Nozieres}
P. Nozi\`eres, \emph{Le probl\`eme \`a  N corps}, Dunod, Paris, (1963).

\bibitem{Abrikosov}
A. A. Abrikosov, L. P. Gorkov, I. E. Dzyaloshinski, \emph{Methods of Quantum Field Theory in Statistical Physics}, Prentice-Hall, (1963).

\bibitem{Baldo03}
M. Baldo, M. Farine, U. Lombardo, E. Saperstein, P. Schuck, M. Zverev,  Eur. Phys. Jour. A \textbf{18} (2003) 17.

\bibitem{Pankratov07}
S. Pankratov, M. Baldo, U. Lombardo, E. Saperstein, M. Zverev,  Phys. At. Nucl. \textbf{70} (2007) 658.

\bibitem{Barranco04}
F. Barranco, R. A. Broglia, G.  Col{\'o}, G. Gori, E. Vigezzi, P. F. Bortignon,  Eur. Phys. Jour. A \textbf{21} (2004) 57.

\bibitem{Terasaki02}
J. Terasaki, F. Barranco, R. A. Broglia, E. Vigezzi, P. F. Bortignon, Nucl. Phys. A \textbf{697},
 (2002) 127.

\bibitem{Giovanardi02}
N. Giovanardi, F. Barranco, R. A. Broglia, E. Vigezzi, Phys. Rev. C \textbf{65} (2002) 041304.

\bibitem{Gori05}
G. Gori, F. Ramponi, F. Barranco, P. F. Bortignon, R. A. Broglia, G. Col\'o, and E. Vigezzi, Phys. Rev. C \textbf{72}, (2005) 011302.

\bibitem{Pastore08}
A. Pastore, F. Barranco, R. A. Broglia, E. Vigezzi, Phys. Rev. C \textbf{78} (2008) 024315.

\bibitem{Duguet07}
T. Duguet, T. Lesinski, Eur. Phys. Jour. ST \textbf{156} (2008) 207.

\bibitem{lesinski07a}
T. Lesinski, Ph.D. thesis, Universit\'e Lyon 1 (2008), 141-2008.

\bibitem{Lesinski08}
T. Lesinski, T. Duguet, K. Bennaceur, J. Meyer, in preparation.

\bibitem{Bogner06b}
S. K. Bogner, R. J. Furnstahl, S. Ramanan, A. Schwenk, Nucl. Phys. A \textbf{773} (2006) 203.

\bibitem{Duguet04}
T. Duguet, Phys. Rev. C \textbf{69} (2004) 054317.

\bibitem{chabanat98}
E. Chabanat, P. Bonche, P. Haensel, J. Meyer, R. Schaeffer, Nucl. Phys. A \textbf{635} (1998) 231.

\bibitem{KaiAchimTD}
K. Hebeler, T. Duguet, T. Lesinski, A. Schwenk, in preparation.

\bibitem{lesinski08b}
T. Lesinski, BSLHFB code: HFB computer code
                  in a spherical Bessel functions basis, 2008, unpublished.

\bibitem{Duguet01}
T. Duguet, P. Bonche, P.-H. Heenen, J. Meyer, Phys. Rev. C \textbf{65} (2001) 014310.

\bibitem{Stoitsov07}
M. Stoitsov, J. Dobaczewski, R. Kirchner, W. Nazarewicz, J. Terasaki, Phys. Rev. C \textbf{76} (2007) 014308.

\bibitem{Lesinski07}
T. Lesinski, M. Bender, K. Bennaceur, T. Duguet, J. Meyer, Phys. Rev. C \textbf{76} (2007) 014312.

\bibitem{Nemirovsky62}
P.~E. Nemirovsky, Y.~V. Adamchuk, Nucl. Phys. \textbf{39} (1962) 551.

\bibitem{lesinski06a}
T. Lesinski, K. Bennaceur, T. Duguet, J. Meyer, Phys. Rev. C \textbf{74} (2006) 044315.

\bibitem{Anguiano01}
M. Anguiano, J. L. Egido, L. M. Robledo, Nucl. Phys. A \textbf{683} (2001) 227.

\bibitem{Goriely06}
S. Goriely, M. Samyn, J. M. Pearson, Nucl. Phys. A \textbf{773} (2006) 279.

\bibitem{yamagami08a}
M. Yamagami, Y. R. Shimizu, Phys. Rev. C \textbf{77} (2008) 064319.

\bibitem{Bogner05}
S. K. Bogner, A. Schwenk, R. J. Furnstahl, A. Nogga, Nucl. Phys. A \textbf{763} (2005) 59.

\bibitem{coraggio03a}
L. Coraggio, N. Itaco, A. Covello, A. Gargano, T. T. S. Kuo, Phys. Rev. \textbf{C68} (2003) 034320.

\bibitem{roth06a}
R. Roth, P. Papakonstatinou, N. Paar, H. Hergert, T. Neff, H. Feldmeier, Phys. Rev. \textbf{C73} (2006) 044312.

\bibitem{brown98a}
B. A. Brown, Phys. Rev. \textbf{C58} (1998) 220.

\bibitem{Barranco05}
F. Barranco, P. F. Bortignon, R. A. Broglia, G. Col\'o, P. Schuck, E. Vigezzi, X. Vi\~nas, Phys. Rev. C \textbf{72} (2005) 054314.

\end{thebibliography}
\end{document}